\documentclass[twocolumn,superscriptaddress,aps,prb]{revtex4-2}
\usepackage[utf8]{inputenc}
\usepackage[T1]{fontenc}
\usepackage{newtxtext,newtxmath}
\usepackage{graphicx}

\usepackage{hyperref}
\usepackage{bm}
\usepackage{microtype}
\usepackage{xcolor}
\usepackage[caption=false]{subfig}

\begin{document}
\title{Topological Hall effect induced by classical large-spin background: $su(2)$ path-integral approach}
\author{Kaushal K. Kesharpu}
\email{kesharpu@theor.jinr.ru}
\affiliation{Bogoliubov Laboratory of Theoretical Physics, Joint Institute for Nuclear Research, Dubna,Moscow Region 141980, Russia}

\author{Evgenii A. Kochetov}
\affiliation{Bogoliubov Laboratory of Theoretical Physics, Joint Institute for Nuclear Research, Dubna,Moscow Region 141980, Russia}

\author{Alvaro Ferraz}
\affiliation{International Institute of Physics - UFRN, Department of Experimental and Theoretical Physics - UFRN, Natal 59078-970, Brazil}
\date{\today}

\begin{abstract}
The $su(2)$ coherent-state path-integral technique is employed to study lattice electrons strongly coupled to a quantum spin background. In the large-spin limit it is replaced by its classical counterpart that breaks the time-reversal symmetry. The fermions propagating through a classical large-spin texture may then exhibit the topological Hall effect which arises even for a zero scalar spin chirality of the underlying spin background.
\end{abstract}

\maketitle

\section{Introduction}
The topological Hall effect (THE) may be viewed as arising from the  electron hopping in a classical spin background that breaks time-reversal symmetry. This results in the anomalous Hall conductivity in the absence of any externally applied magnetic field. The THE attracts attention from both physics and engineering communities due to the novel physics it contains and due to the potential applications to electronics and spintronics.
The simplest examples come from a mean-field treatment of the Kondo-Lattice interaction (the double exchange model) of the itinerant electrons and the classical localized spins that form non-coplanar spin textures~\cite{yeBerryPhaseTheory1999,chunMagnetotransportManganitesRole2000,ohgushiSpinAnisotropyQuantum2000,yiSkyrmionsAnomalousHall2009}.

Specifically, one starts with the simplest lattice Kondo-type model of the tight-binding electrons strongly coupled to the localized spins $\hat{\mathbf{S}}_i$ described by the Hamiltonian
\begin{equation}
  \label{eq:ham-kondo}
  \begin{aligned}
    H=-\sum_{ij\sigma}&(t_{ij}+\frac{3J}{4}\delta_{ij})c^{\dagger}_{i\sigma}c_{j\sigma}\\
    &+J\sum_i \hat{\mathbf{S}}_i\cdot(c^{\dagger}_{i\sigma}\vec{\sigma}_{\sigma\sigma'}c_{i\sigma'}).
  \end{aligned}
\end{equation}
Here $c^{\dagger}_{i\sigma}$ creates an electron with the spin $\sigma$ on site $i$. $J>0$ stands for the exchange coupling constant. $\vec{\sigma}$ is the vector of the \emph{Pauli} spin matrices. The external magnetic moments $\hat{\mathbf{S}}_i$ are the generators of the $su(2)$ algebra in the lowest $(s=1/2)$ representation. The hopping term is modified by adding an extra $J$ dependent term to guarantee a finite $J\to +\infty$ limit~\cite{ivantsov_2022_AnnalsofPhysics}. Within a mean-field treatment the spatial spin structure is encoded in $<\hat{\mathbf{S}}_i> = S \cdot \vec{n}_i$. Here $S$ is a localized spin magnitude,
whereas a classical static vector $\vec{n}_i$ determines a varying direction of the localized spin.

The underlying topological spin structures can be composed of multiple spin density waves. The scalar spin chirality in the ground state due to spins of the \emph{l}, \emph{m}, and \emph{n}-th site is found by the triple product $\langle \vec{S}_l\cdot \vec{S}_m\times \vec{S}_n \rangle$. For non-zero spin chirality, the time-reversal and the parity symmetries are broken. Hence, when conduction electrons propagate through such a spin texture, due to accumulation of \emph{Berry} phase, the system may show THE~\cite{martinItinerantElectronDrivenChiral2008}. Typical postulated spin textures are chiral stripes, spiral spin configurations, skyrmion spin structures~\cite{hayamiChargeDensityWaves2021,hayamiRectangularSquareSkyrmion2022}. In particular the experimentally observed  skyrmion lattice can be viewed as a lattice of topologically stable knots in the underlying spin structure~\cite{neubauerTopologicalHallEffect2009,hanSkyrmionLatticeTwodimensional2010}.

Away from the mean-field treatment one should consider operators $\hat{\mathbf{S}}_i$ for $S>1/2$ as being the generators of the higher spin-$S$ representation, which poses a technical problem. On the other hand, the $su(2)$ coherent state (CS) path integral incorporates $S$ just as a parameter~\cite{kochetovSUCoherentState1995}. The CS integral is entirely determined by the $su(2)$ algebra commutation relations that do not depend on a chosen representation. It therefore seems appropriate to alternatively  treat the problem in terms of the CS path-integral representation for the partition function. In this way one avoids explicit matrix representations of the high-spin quantum  generators, and instead deals with a compact and simple $su(2)$ path integral for the partition function. Of course by postulating the underlying spin texture as a static site-dependent field we arrive again at this stage (after the limit $J \to\infty$ is taken) in a new mean-field approximation.

In our previous work~\cite{ferrazFractionalizationStronglyCorrelated2022}, we proposed such a theory to treat the THE based on the lowest $s=1/2$ representation of the $su(2)$ algebra. The aim of this present work is twofold. First, we generalize our approach to study large-spin classical textures. This is important as that seems to be the case in many theoretical and experimental studies~\cite{martinItinerantElectronDrivenChiral2008,buessenQuantumSpinLiquids2018,chamorroFrustratedSpinOne2018}.  Second, until recently, most investigations have been focused on magnetic states with a finite spin chirality. We show that the THE does not necessarily require a nonzero scalar spin chirality. Although in the physical community it is widely believed that THE is absent for zero spin chirality~\cite{nagaosaTopologicalPropertiesDynamics2013}, recent experiments have shown otherwise~\cite{afsharSpinSpiralTopological2021,mendezCompetingMagneticStates2015,ghimireCompetingMagneticPhases2020,gongLargeTopologicalHall2021,wangFieldinducedTopologicalHall2021}.

Recently, the THE in magnets having a trivial magnetic structure has been attracting broad interest. For example, a large magnitude of the crystal THE is proposed in a room-temperature collinear antiferromagnet RuO$_{2}$. Namely, if the crystal symmetry is low enough, the THE can emerge even when the background spin texture is trivial~\cite{smejkalCrystalTimereversalSymmetry2020}.
It is possible to have the THE in a noncollinear antiferromagnet with zero net magnetization. In this respect, the topological order indeed survives in the magnetic material Mn$_{3}$Ir to very high temperatures~\cite{chenAnomalousHallEffect2014}. The THE is also shown to emerge in non-collinear spiral spin textures in a $2d$ magnetic Rashba model \cite{luxChiralHallEffect2020,mochidaSkewScatteringMagnetic2022}. In the present work, we show that spin-orbit coupling is not truly a necessary ingredient for the realization of the THE exhibiting a trivial magnetic structure. It can instead be driven by time-reversal symmetry breaking in a system of strongly correlated electrons that exhibits zero spin chirality.

\section{Theoretical framework}\label{sec:theor-fram}
A low-energy effective theory to describe the electrons coupled to a spin-\emph{S} background can be derived under the requirement that, the spin background should affect the fermion hopping in such a way that the $SU(2)$ global symmetry remains intact. The electron hopping is then affected by the $su(2)$ CS overlap factor. This is analogous to the \emph{Peierls} factor arising due to an external magnetic field. It is frequently referred to as the \emph{vector potential} generated by a noncollinear spin texture~\cite{nagaosaTopologicalPropertiesDynamics2013}. Physically, one can view it as an \emph{fictitious} magnetic field that produces a flux through an elementary plaquette. The precise meaning of that emergent \emph{artificial gauge field} is as follows. It is generated by the $U(1)$ local connection one-form of the spin $U(1)$ complex line bundle. Such a construction provides a covariant (geometric) quantization of  a spin~\cite{stoneSupersymmetryQuantumMechanics1989}. In this approach the underlying base space appears as a classical spin phase space --- a two sphere $S^{2}$. It can be thought of as a complex projective space $CP^{1}$, endowed with a set of local coordinates $(z,\bar{z})$. Quantum spin is then represented as the sections $|z\rangle$ of the principle (monopole) line bundle $P(CP^1, U(1))$. The local connections of this bundle read $a^{(0)}=i\langle z|d|z\rangle$,  with $d$ standing for an exterior derivative.

This approach proves effective in studying strongly correlated electrons. This can be seen as follows. At infinitely large \emph{Kondo} coupling $J$, Eq.~(\ref{eq:ham-kondo}) goes over into the $U=\infty$ \emph{Hubbard} model Hamiltonian~\cite{ivantsov_2022_AnnalsofPhysics}:
\begin{equation}
  \label{eq:large-U-ham}
  H_{U=\infty}=-\sum_{ij\sigma}t_{ij}\tilde c_{i\sigma}^{\dagger}\tilde c_{j\sigma}.
\end{equation}
The constrained electron operator $\tilde c_{i\sigma}=c_{i\sigma}(1-n_{i\bar\sigma})$ --- where $n_{i\sigma}=c^{\dagger}_{i\sigma}c_{i\sigma}$ is the number operator --- can be dynamically (in the effective action) factorized into the spinless charged fermionic $f_i$ fields and the spinfull bosonic $z_i$ fields~\cite{ferrazFractionalizationStronglyCorrelated2022}. Inasmuch as $f_i^2\equiv 0$, the local no double occupancy constraint that incorporates the strong electron correlations is rigorously implemented in this representation.

The high-spin extension of the present theory can be obtained by simply generalizing the CS in the fundamental $S=1/2$ representation for a given spin-\emph{S}:
\begin{equation}
  \label{eq:CS-S-rep}
  |z\rangle= \left(1+|z|^2\right)^{-S}e^{z\hat S^{-}}|S\rangle,
\end{equation}
where $|S\rangle$ represents the spin-$S$ highest weight $su(2)$ state, and the operator $\hat S^{-}$ denotes a conventional spin lowering operator. As a result the \emph{S}-dependent partition function takes on the form
\begin{equation}
  \label{eq:part-fun}
Z=\int D\mu (z,f) \exp\mathcal{A}.
\end{equation}
Here the measure $D\mu(z,f)$ is defined as:
\begin{equation}
  \label{eq:measure}
  D\mu (z,f)=\prod_{i,t}\frac{2S}{2\pi i}\frac{d\bar z_i(t)dz_i(t)}{\left(1+|z_i|^2\right)^2}\, d\bar f_i(t)d f_i(t).
\end{equation}
In Eq.~(\ref{eq:measure}) $z_i$ is a complex number that keeps track of the spin degrees of freedom, while $f_i $ is a \emph{Grassmann} variable that describes the charge degrees of freedom.

The effective action $\mathcal{A}$ in Eq.~(\ref{eq:part-fun}) is defined as:
\begin{equation}
  \label{eq:effective-act}
  \mathcal{A} = \sum\limits_{i} \int\limits_{0}^{\beta} \left[ i a_{i}^{(0)}- \bar{f}_{i} \left( \partial_{t} + i a_{i}^{(0)} \right)f_{i}\right]dt - \int\limits_{0}^{\beta} H dt.
\end{equation}
It involves the $u(1)$-valued connection one-form of the magnetic monopole bundle that can formally be interpreted as a \emph{spin kinetic} term
\begin{equation}
  \label{eq:spin-kinetic}
   ia^{(0)}=-\langle z|\partial_t|z\rangle=S \: \frac{\dot{\bar{z}} z-\bar{z}\dot{z}}{1+|z|^2},
\end{equation}
which can be identifies as the Berry connection. The dynamical part of the action can be written as
\begin{equation}
  \label{eq:dynamic-act}
  H = -t \sum\limits_{ij} \bar{f}_{i}f_{j} \mathrm{e}^{i a_{ji}} + H.c. + \mu \sum\limits_{i} \bar{f}_{i}f_{i}.
\end{equation}
Here,
\begin{equation*}
  \label{eq:phase-aji}
  \begin{aligned}
    &a_{ij} = -i \log \langle z_{i}|z_{j} \rangle,
    \langle z_{i}|z_{j} \rangle =\frac{\left(1+\overline{z} _{i}z_{j} \right)^{2S}}{\left( 1+|z_{j}|^{2} \right)^S \left(  1+|z_{i}|^{2}\right)^S}.
  \end{aligned}
\end{equation*}

The hopping term  in Eq.~(\ref{eq:dynamic-act}) is affected by the $su(2)$ coherent-state overlap factor in the $S$- representation.
The modulus of this factor
\begin{equation}
\left|\langle z_i|z_j\rangle \right|= \left[\frac{(1+\bar z_i z_j)(1+\bar z_j z_i)}{(1+|z_i|^2)(1+|z_j|^2)}\right]^S=:\rho_{ij}^S.
\end{equation}
Here $\rho_{ij}\in [0,1]$. The hopping probability is reduced by the presence of a factor $\rho_{ij}^S$  which is proportional to the overlap of the spin wave functions on neighboring sites. In the limiting FM case ($z_i=z_j$) those wave functions are identical and $\rho=1$ at any spin value $S$. There is no overlap in the AFM case $(z_j=-1/z_i)$ at any $S$, hence $\rho=0$. The hopping creates changes in the spin configuration, unless the spin polarization is uniform. The growing $S$ inhibits the hopping by reducing the spin-wave functions overlap. As a result, away from the FM case as $\rho_{ij}<1$, and, eventually,  with increasing $S\to\infty$ the $\rho_{ij}^S \to 0$. This is the expected physical behavior in such a regime ~\cite{shankar_1990_NuclearPhysicsB}.

The passage from Eq.~(\ref{eq:ham-kondo}) to Eq.~(\ref{eq:dynamic-act}) implies that we take the limit $J/t\to\infty$ keeping at the same time the $su(2)$ spin generators $\hat{\mathbf{S}}_i$ fixed. 
In this limit the quantum local spins are fully screened and cannot be replaced by fixed classical values. The spin classical limit implies instead  that we replace from the very beginning 
$\hat{\bf S}_i$ in Eq. (\ref{eq:ham-kondo}) by the $c$-numbers $S \langle \vec{n}_i\rangle$. This fully ignores quantum \emph{Kondo screening} \footnote{This model is applicable for small as well as large values of the $S$. The only physical consequence is the decrease in the hopping probability for increasing  as explained in the previous reply. We start with the model from Eq. (\ref{eq:ham-kondo}) in the limit $J/t \to \infty$ the $su(2)$ representation being fixed. This limit enforces the constraint of no double occupancy and must be taken prior to any mean-field treatment. Fixing of the  representation fixes $S$ as discussed in Eq. (\ref{eq:CS-S-rep}). We thus restrict ourselves with finite $S \geq 1/2$.}.

Under a \emph{global} $SU(2)$ rotation
\begin{equation}
  \label{eq:su2-rotation}
  z_{i} \to \frac{u z_{i} + v}{- \bar{v} z_{i}+\bar{u}},
\end{equation}
the phase will be:
\begin{equation}
  \label{eq:su-2-phase}
  a_{i}^{(0)} \to a^{(0)}_{i} - \partial_{t} \theta_{i}, \quad a_{ij} \to a_{ij} + \theta_{j} - \theta_{i},
\end{equation}
where,
\begin{equation}
  \label{eq:theta-su2}
  \theta_{i} = i S \log \left( \frac{v \bar{z}_{i} + u}{\bar{v} z_{i} + \bar{u}} \right); \quad
  \begin{pmatrix}
    u &v\\
    -\bar{v} &\bar{u}
  \end{pmatrix}
  \in SU(2).
\end{equation}
One can clearly observe that, the effective action $\mathcal{A}$ remains invariant under $SU(2)$ transformations, provided the fermionic operators are $U(1)$ transformed as:
\begin{equation}
  \label{eq:fermion-u1-tran}
  f_{i} \to e^{i \theta_{i}}f_{i}.
\end{equation}
A \emph{flux} through a plaquette $\sum_{\text{plaq}}a_{ij}$ generated by the $SU(2)$ transformation remains invariant under Eq.~(\ref{eq:su-2-phase}).

By definition the phase $a_{ji}$ in Eq.~(\ref{eq:dynamic-act}) is a complex valued function. Hence, the real and imaginary part of $a_{ij}$ are:
\begin{equation}
  \label{eq:aij-real-imag}
  a_{ji} = \phi_{ji}+i\chi_{ji}; \quad \bar{\phi}_{ji}=\phi_{ji}; \quad \bar{\chi}_{ji}=\chi_{ji}.
\end{equation}
The $\phi_{ji}$ and $\chi_{ji}$ are defined as
\footnote{The CS symbols of the spin operators, $S_{cs}:=\langle z|\hat S|z \rangle$, are:
\begin{equation*}
  \label{eq:CS-S-symb}
  S^{+}_{cs}=\frac{2Sz}{1+|z|^2}, \: S^{-}_{cs}=\frac{2S\bar z}{1+|z|^2}, \:S^z_{cs}= S\frac{1-|z|^2}{1+|z|^2}.
\end{equation*}
There is a one-to-one correspondence between the $su(2)$ generators and their CS symbols~\cite{berezinIntroductionSuperanalysis1987}.}:
\begin{equation}
  \label{eq:phi-chi-ji}
  \begin{aligned}
    &\phi_{ji} &&=iS \log \frac{1 + \bar{z}_{i} z_{j}}{1+ \bar{z}_{j} z_{i}}\\
    &          &&=iS \log \frac{\left( S+S_{i}^{z} \right)\left( S+S_{j}^{z} \right)+S_{i}^{-} S_{j}^{+}}{\left( S+S_{i}^{z} \right)\left( S+S_{j}^{z} \right)+S_{j}^{-} S_{i}^{+}};\\
    &\chi_{ji} &&= -S \log \frac{\left( 1+\bar{z}_{i}z_{j} \right)\left( 1+\bar{z}_{j}z_{i} \right)}{\left( 1+|z_{i}|^{2} \right) \left( 1+|z_{j}|^{2} \right)}\\
    &   &&=- S \log \left( \frac{\vec{S}_{i} \cdot \vec{S}_{j}}{2S^{2}} + \frac{1}{2} \right).
  \end{aligned}
\end{equation}
Using Eq.~(\ref{eq:phi-chi-ji}) it can be checked that, under a $SU(2)$ global rotation $\chi_{ji}$ remains intact, however, $\phi_{ji}$ transforms as:
\begin{equation}
  \label{eq:phi-ji-trans}
  \phi_{ji}  \to \phi_{ji} + \theta_{i}  - \theta_{j}.
\end{equation}
This transformation appears as a gauge fixing by choosing a specific rotational covariant frame. The dynamical fluxes do not depend on that choice.

The potentials $\phi_i^{(0)}:=ia^{(0)}_i$ and $\phi_{ji}$ formally  remind those gauge fields that define a compact $U(1)$ lattice gauge theory. This is due to the fact that both theories are formulated as $U(1)$ complex line bundles. The gauge potentials (local connections in these bundles) in both theories transform formally in the same way under a change in the local trivialization~\cite{nakaharaGeometryTopologyPhysics2018}. In our case, the different trivializing coverings of the spin base manifold $S^2$ are related to each other through the global $SU(2)$ rotations --- Eq.~(\ref{eq:su2-rotation}).

\section{Topology}\label{sec:topology}
Using Eq.~(\ref{eq:phi-chi-ji}) the Hamiltonian can be explicitly written as:
\begin{equation}
  \label{eq:gen-ham}
  \begin{aligned}
    H = -t \sum\limits_{\left\langle i,j \right\rangle} \bar{f}_{i} f_{j} \mathrm{e}^{i \phi_{ji}} \left( \frac{\vec{S}_{i} \cdot \vec{S_{j}}}{2 S^{2}} + \frac{1}{2} \right)^{S} + \mu \sum\limits_{i} \bar{f}_{i} f_{i}.
  \end{aligned}
\end{equation}
Physically, it represents the interaction between an underlying spin texture and the itinerant spinless fermions. When $S=1/2$ this Hamiltonian reduces to that derived in Ref.~\cite{ferrazFractionalizationStronglyCorrelated2022}. Here we generalize our approach to large values of spin $S \gg 1/2$. After all the classical treatment of underlying spin structure is more justifiable when $S$ is larger than the electronic spin.

We take a two band system having opposite \emph{Chern} number. Physically, one can consider a bipartite \emph{2D} lattice \emph{L}, consisting of two sub-lattice \emph{A} and \emph{B}; $L=A \oplus B$. If on the sub-lattice \emph{A} the charge and spin degrees of freedom are $f_{i}$ and $z_{i}$, respectively, then for convenience on the sub-lattice \emph{B} they are defined as:
\begin{equation}
  \label{eq:sub-lat-B-dof}
   f_i \to f_{i} \mathrm{e}^{i \theta_{i}^{(0)}}, \quad z_{i} \to -\frac{1}{\bar{z}_{i}}; \qquad i \in B.
\end{equation}
Here $\theta_{i}^{(0)} \equiv \theta_{i}|_{u=0,v=1}$ in Eq.~(\ref{eq:theta-su2}). Under these transformations the $\chi_{ji}$ remains unchanged while the $\phi_{ji} \to \phi_{ij} + \theta_{j}^{(0)} - \theta_{i}^{(0)}$. Also the \emph{CS} image of the on-site electron spin operators changes sign, $\vec{S}_{i} \to - \vec{S}_{i}$. As $\phi_{ji}=-\phi_{ij}$ we see that the phase of the hopping factor corresponding to the \emph{A} and \emph{B} sub-lattice are opposite in sign. This means that the dynamical fluxes piercing through elementary plaquettes that compose a unit cell are opposite in sign. Fixing them as classical \emph{c}-numbers breaks time-reversal symmetry. Indeed they are analogous to the homogeneous \emph{c}-valued phase factors in the \emph{NNN} hopping, $t_{2} \to t_{2} \mathrm{e}^{\pm i \phi}$, introduced by Haldane~\cite{haldaneModelQuantumHall1988}. Hence, as in the Haldane model, one can expect a non zero \emph{Hall} effect emerging due to the resulting time reversal symmetry breaking. It should be noted that, the time reversal symmetry in the Haldane model was broken by inserting non-zero local fluxes; with the summation of those fluxes over the unit cell being zero. In contrast, in our model Eq.~(\ref{eq:dynamic-act}) we simply break time reversal symmetry by fixing the spin connection term $a_{ij}$ as a classical static quantity. In fact fixing the $a_{ij}$ may result in breaking time reversal even for zero spin chirality.

\subsection{An example: Conical spin configuration}\label{sec:conic-spin-conf}
Below we take a simple spin configuration to illustrate our approach. Interestingly this spin structure shows non-trivial topological effect, although its scalar spin chirality is zero. The spin-\emph{S} field is defined as:
\begin{equation}
  \label{eq:con-spin}
  \vec{S}_{i} = \left( S_{i}^{x}, S_{i}^{y}, S_{i}^{z} \right) \equiv \left( \epsilon \cos \vec{q} \cdot \vec{r}_{i}, \epsilon \sin \vec{q} \cdot \vec{r}_{i}, \sqrt{S^{2}-\epsilon^{2}} \right).
\end{equation}
Here $\vec{S}_{i}$ is the spin at \emph{i}-th site, $\vec{q}$ is the spin modulation vector, $\vec{r}_{i}$ is the position vector, with $S$ being some arbitrary constant spin value, and $\epsilon$ some small parameter satisfying $\epsilon \ll S$. Physically, it represents the precession of spin $S$ around \emph{z}-axis with a constant azimuthal angle $\pi/2-\arctan \left( \sqrt{S^{2}-\epsilon^{2}}/\epsilon \right)$. This formalism is applicable to a thin film or to a surface of a  multiferroic single-crystal assuming a constant $|\vec{S}_{i}|=S$~\cite{chenMajoranaEdgeStates2015}. Keeping the terms only up to second order in $\epsilon$, the hopping terms contributions  of the Hamiltonian --- Eq.~(\ref{eq:gen-ham}) ---  only for the sub-lattice \emph{A} are:
\begin{equation}
  \label{eq:sub-a-term}
  \begin{aligned}
    \mathrm{e}^{-\chi_{ij}} & \approx 1 - \frac{\epsilon^{2}}{S} \sin^{2}\frac{\vec{q} \cdot \vec{r}_{ij}}{2},\\
    \mathrm{e}^{i \phi_{ji}} & \approx \exp \left(  i \frac{\epsilon^{2}}{2S} \sin \vec{q} \cdot \vec{r}_{ij} \right).
  \end{aligned}
\end{equation}
Here we define $\vec{r}_{ij}=\vec{r}_{i}-\vec{r}_{j}$. At the sub-lattice \emph{B} the $\chi_{ij}$ remains the same, but the $\phi_{ji} \to -\phi_{ji}|_{i,j \in A} + 2S \vec{q} \cdot  \vec{r}_{ij}$. Therefore
\begin{equation}
  \label{eq:sub-b-term}
  \begin{aligned}
    \mathrm{e}^{-\chi_{ij}} & \approx 1 - \frac{\epsilon^{2}}{S} \sin^{2}\frac{\vec{q} \cdot \vec{r}_{ij}}{2},\\
    \mathrm{e}^{i \phi_{ji}} & \approx \exp \left(-i \frac{\epsilon^{2}}{2S} \sin \vec{q} \cdot \vec{r}_{ij} + i 2S \vec{q} \cdot \vec{r}_{ij} \right).
  \end{aligned}
\end{equation} Siimilarly, for inter-lattice hoppings ($i \in A, j \in B$):
\begin{equation}
  \label{eq:sub-ab-term}
  \begin{aligned}
    \mathrm{e}^{-\chi_{ij}} & \approx \left( \frac{\epsilon}{S} \sin\frac{\vec{q} \cdot \vec{r}_{ij}}{2} \right)^{2S},\\
    \mathrm{e}^{i \phi_{ji}} & \approx e^{iS \left( \vec{q} \cdot \vec{r}_{ij}  - \pi\right)}.
  \end{aligned}
\end{equation}
The total Hamiltonian is found by substituting Eq.~(\ref{eq:sub-a-term}), (\ref{eq:sub-b-term}) and (\ref{eq:sub-ab-term}) in Eq.~(\ref{eq:gen-ham}):
\begin{equation}
  \label{eq:con-spin-ham}
  \begin{aligned}
    H =
      & -t_{2} \sum\limits_{i,j \in A} \bar{f}_{i} f_{j} \left( 1 + i \frac{\epsilon^{2}}{S} \sin \frac{\vec{q} \cdot \vec{r}_{ij}}{2} e^{i \vec{q} \cdot \vec{r}_{ij}/2}\right)\mathrm{e}^{- i S \vec{q} \cdot \vec{r}_{ij}}\\
      & -t_{2} \sum\limits_{i,j \in B} \bar{f}_{i} f_{j} \left( 1 - i \frac{\epsilon^{2}}{S} \sin \frac{\vec{q} \cdot \vec{r}_{ij}}{2} e^{-i \vec{q} \cdot \vec{r}_{ij}/2}\right)\mathrm{e}^{i S \vec{q} \cdot \vec{r}_{ij}}\\
      & -t_{1} \left( \frac{\epsilon}{S} \right)^{2S} \sum\limits_{\substack{i \in A \\ j \in B}} \bar{f}_{i} f_{j} \left| \sin \frac{\vec{q} \cdot \vec{r}_{ij}}{2} \right|^{2S} + H.c.
  \end{aligned}
\end{equation}
It should be stressed that, the $\vec{r}_{ij}$ in the first and in the second terms are different from the $\vec{r}_{ij}$ in the third term. In Eq.~(\ref{eq:con-spin-ham}) the third term on the right hand side is real and it corresponds to  \emph{NN} hopping contributions. The first and the second term correspond to the \emph{NNN} hopping processes and they are complex valued; in fact they are conjugate of each other, which is a necessary condition for time reversal symmetry breaking.

In momentum space the Hamiltonian is written as:
\begin{equation}
  \label{eq:con-spin-ham-k}
  H = \sum\limits_{\vec{k}} \bar{\psi}_{\vec{k}} \mathcal{H}(\vec{k}) \psi_{\vec{k}}.
\end{equation}
Here, $\vec{k}$ is the wave vector whose values lie only in the first \emph{Brillouin zone}. The matrix $\bar{\psi}_{\vec{k}} = \left[ \bar{f}_{\vec{k},A} \:,\: \bar{f}_{\vec{k},B} \right]$ contains the annihilation operators of the \emph{$\vec{k}$}-th momentum on the \emph{A} and \emph{B} sub-lattices. The kernel $\mathcal{H}(\vec{k})$ is
\footnote{$H(\vec{k})$ is a $2 \times 2$ matrix. In terms of \emph{Pauli matrices} it is represented as
  \begin{equation*}
    \label{eq:pauli-mat}
    \begin{aligned}
      &\mathcal{H}(\vec{k}) = \mathcal{H}_{0} \mathcal{I} + \mathcal{H}_x (\vec{k}) \sigma_{x} + \mathcal{H}_y (\vec{k}) \sigma_{y} +\mathcal{H}_z (\vec{k}) \sigma_{z},\\
      &\text{where,}\\
      &\mathcal{H}_{0}(\vec{k}) = \frac{H_{i,j \in A} + H_{i,j \in B}}{2}, \mathcal{H}_{x}(\vec{k}) = \Re \left[ H_{i \in A, j \in B} \right],\\
      &\mathcal{H}_{z}(\vec{k})=\frac{H_{i,j \in A} - H_{i,j \in B}}{2},  \mathcal{H}_{y}(\vec{k}) = \Im \left[ H_{i \in A, j \in B} \right].
    \end{aligned}
  \end{equation*}
  Here, $\mathcal{I}$ is the $2 \times 2$ unit matrix; $\sigma_{x}$, $\sigma_{y}$, and $\sigma_{z}$ are the Pauli matrices.
}
:
\begin{widetext}
% \vspace{\columnsep}
\begin{equation}
  \label{eq:kern-hk-con-spin}
  \begin{aligned}
    \mathcal{H}(\vec{k})=&\left\{t_{2} \sum\limits_{n}\cos \vec{k} \cdot \vec{b}_{n} \left[ \cos S\vec{q} \cdot \vec{b}_{n} - \frac{\epsilon^{2}}{S} \sin \frac{\vec{q} \cdot \vec{b}_{n}}{2} \sin \frac{\left( 2S-1\right)\vec{q} \cdot \vec{b}_{n}}{2}\right] \right\}\cdot \mathcal{I} + \left\{ t_{1} \left( \frac{\epsilon}{S} \right)^{2S} \sum\limits_{n}\cos \vec{k} \cdot \vec{a}_{n} \left|\sin \frac{\vec{q} \cdot \vec{a}_{n}}{2} \right|^{2S} \right\} \cdot \sigma_{x}\\
     & + \left\{ t_{1} \left( \frac{\epsilon}{S} \right)^{2S} \sum\limits_{n}\sin \vec{k} \cdot \vec{a}_{n} \left|\sin \frac{\vec{q} \cdot \vec{a}_{n}}{2} \right|^{2S} \right\} \cdot \sigma_{y}
    + \left\{t_{2} \sum\limits_{n} \sin \vec{k} \cdot \vec{b}_{n} \left[ \sin S\vec{q} \cdot \vec{b}_{n} + \frac{\epsilon^{2}}{S} \sin \frac{\vec{q} \cdot \vec{b}_{n}}{2} \cos \frac{\left( 2S-1 \right)\vec{q} \cdot \vec{b}_{n}}{2}\right] \right\} \cdot \sigma_{z}.
  \end{aligned}
\end{equation}
\vspace{\columnsep}
\end{widetext}
Here $\vec{b}_{n}$ is the \emph{NNN}, and $\vec{a}_{n}$ is the \emph{NN} hopping lattice vectors; $\mathcal{I}$ is the $2 \times 2$ unit matrix; $\sigma_{x}$, $\sigma_{y}$, and $\sigma_{z}$ are the \emph{Pauli matrices}. One can clearly observe that the time reversal symmetry is broken in Eq.~(\ref{eq:kern-hk-con-spin}), as terms corresponding to $\sigma_{z}$ are odd in $k$ ($\propto \sin \vec{k} \cdot \vec{b}_{n}$).

\begin{figure}
     \centering
     \subfloat[][Bipartite honeycomb lattice]{\includegraphics[width=0.22\textwidth]{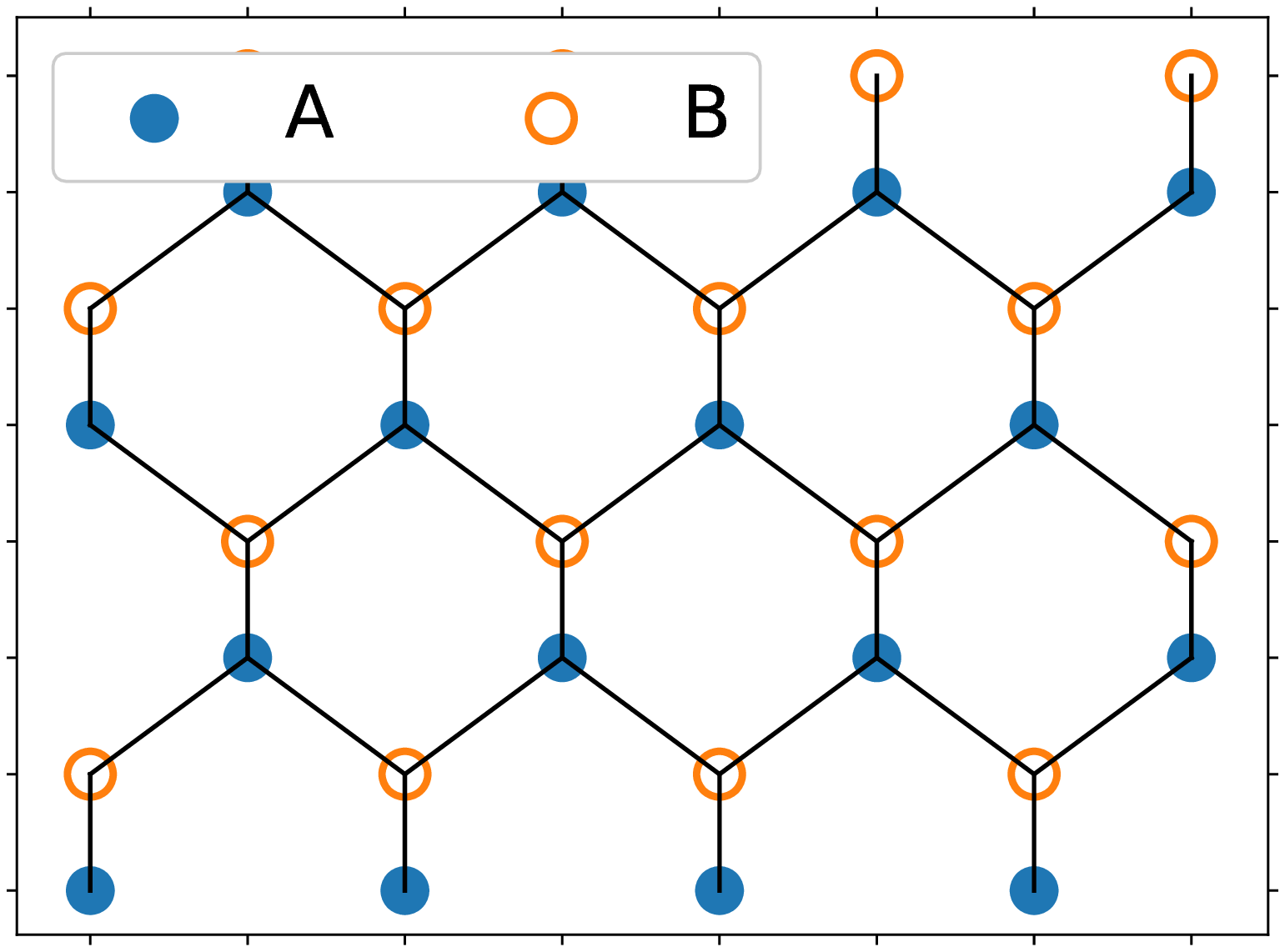}\label{fig:hc-lat-a}}
     \subfloat[][\emph{NN} and \emph{NNN} unit vectors]{\includegraphics[width=0.22\textwidth]{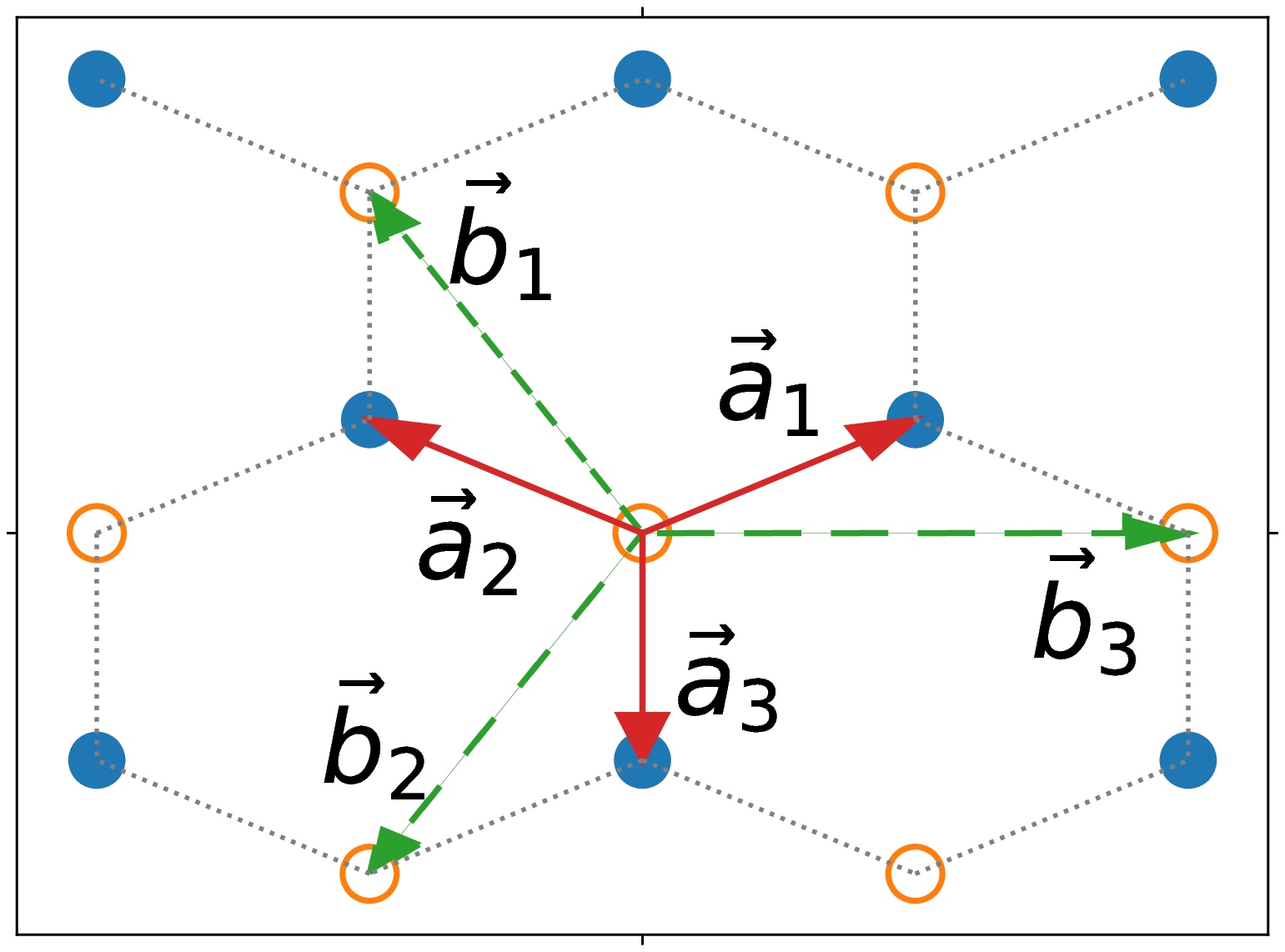}\label{fig:hc-lat-b}}\\
     \subfloat[][Reciprocal lattice]{\includegraphics[width=0.22\textwidth]{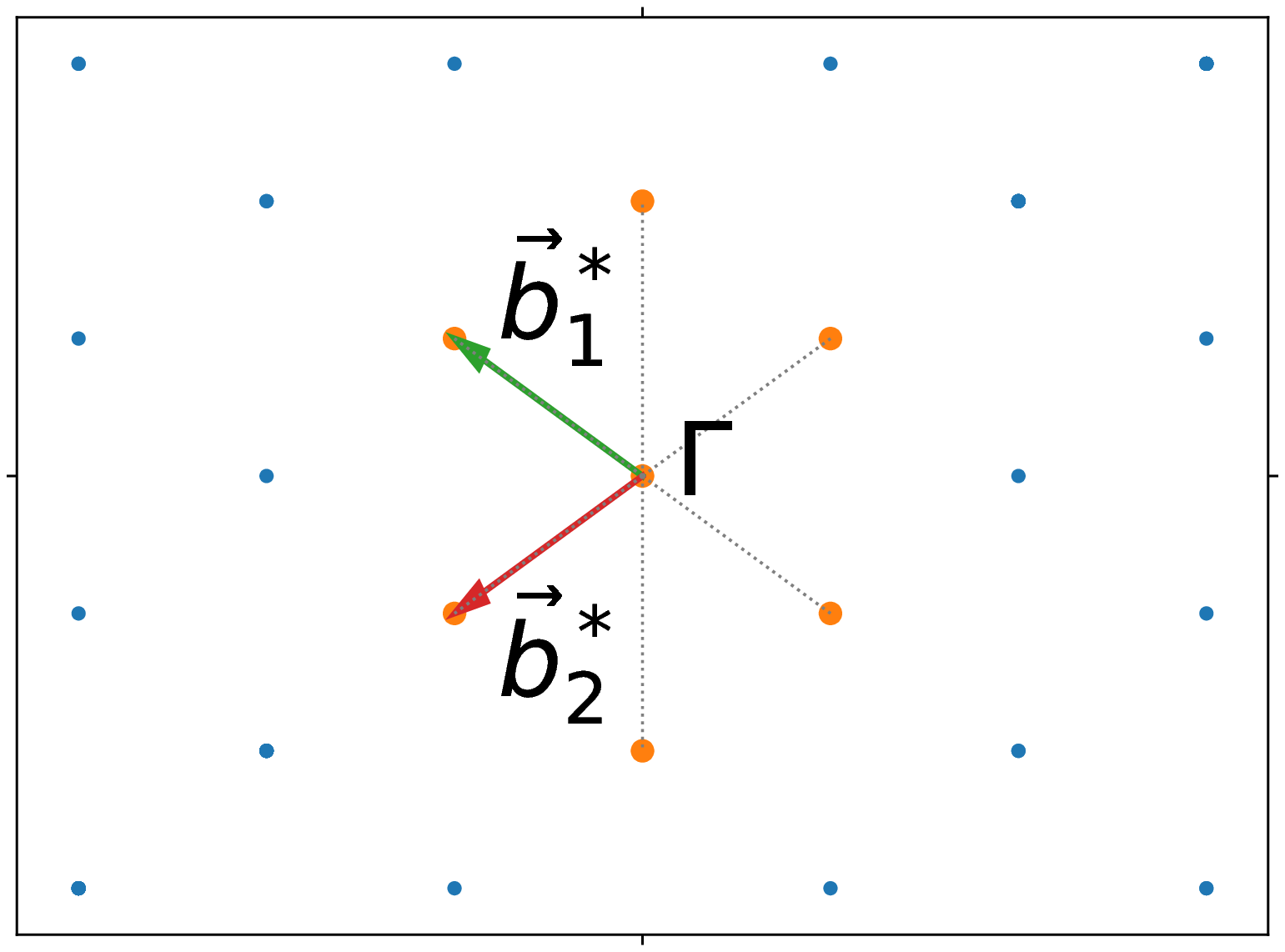}\label{fig:rec-lat}}
     \subfloat[][First \emph{Brillouin} zone]{\includegraphics[width=0.22\textwidth]{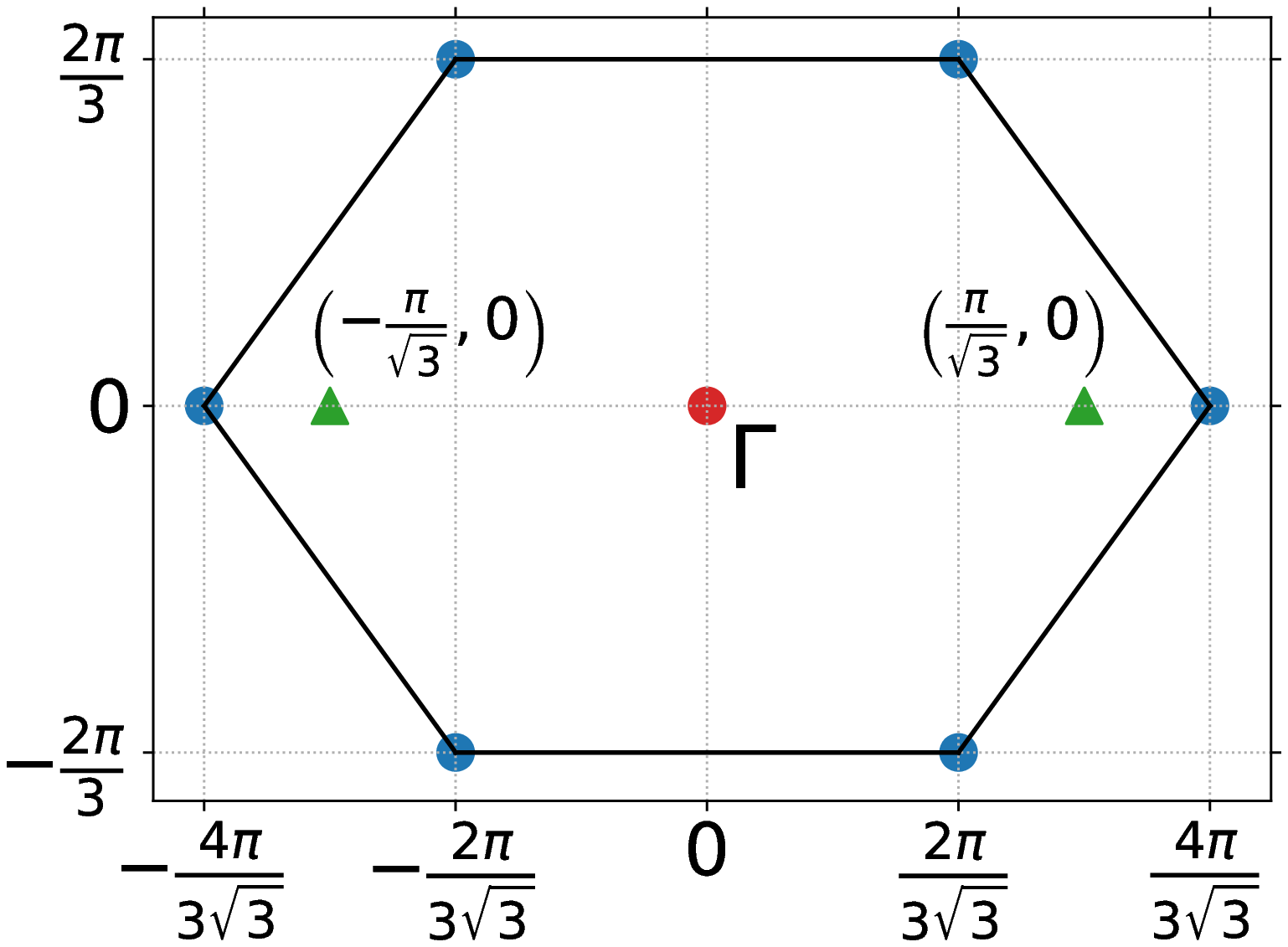}\label{fig:bz}}\\
     \caption{(a) Schematic of the honeycomb bipartite lattice with two types of atoms \emph{A} and \emph{B}. (b) The \emph{NN} vectors: $\vec{a}_{1}$, $\vec{a}_{2}$, and $\vec{a}_{3}$; and the \emph{NNN} vectors: $\vec{b}_{1}$, $\vec{b}_{2}$, and $\vec{b}_{3}$. (c) The reciprocal space of the honeycomb bipartite lattice. The reciprocal unit vectors are: $\vec{b}_{1}^{*}=2 \pi \left( -\frac{1}{\sqrt{3}}, \frac{1}{3} \right)$, $\vec{b}_{2}^{*}=2 \pi \left( -\frac{1}{\sqrt{3}}, -\frac{1}{3} \right)$. (d) The first Brillouin zone of the honeycomb bipartite lattice. The special points (triangle) --- when term $\sim$ $\sigma_{x}$ and $\sim$ $\sigma_{y}$ are zero --- for $\vec q=\left( 2 q/\sqrt{3},0 \right)$ lies at $\vec{K}=\left( \pm \pi/\sqrt{3},0 \right)$.}\label{fig:hc-lat}
\end{figure}
For a bipartite honeycomb lattice the \emph{NN} and the \emph{NNN} lattice vectors are (see Fig.~\ref{fig:hc-lat-b}):
\begin{equation*}
  \label{eq:honey-lattice-vect}
  \begin{aligned}
    &\vec{a}_{1}= \left( \frac{\sqrt{3}d}{2}, \frac{d}{2} \right), \:\quad \vec{a}_{2}= \left(- \frac{\sqrt{3}d}{2}, \frac{d}{2} \right),\:\quad \vec{a}_{3}= \left( 0, d \right),\\
    &\vec{b}_{1}= \left( -\frac{\sqrt{3}d}{2}, \frac{3d}{2} \right), \vec{b}_{2}= \left( -\frac{\sqrt{3}d}{2}, -\frac{3d}{2} \right), \vec{b}_{3}= \left( \sqrt{3}d,0 \right).
  \end{aligned}
\end{equation*}
Here, $d$ is the lattice unit length, and for simplicity we take  $d=1$. These vectors can be substituted in Eq.~(\ref{eq:kern-hk-con-spin}) to get the corresponding Hamiltonian. Two different cases arise for the resulting Hamiltonian: (i) $t_{2}=0$ and $\epsilon \neq 0$, (ii) $t_{2} \neq 0$ and $\epsilon = 0$. For the first case the system is an insulator if the terms corresponding to $\sigma_{x}$ and $\sigma_{y}$ are non zero over the \emph{Brillouin} zone. However, depending on the spin wave vector $q$, there can exist some special values of $k$ for which the terms corresponding to $\sigma_{x}$ and to $\sigma_{y}$ are also zero. At these special points the energy is degenerate. This degeneracy can be lifted by adding the \emph{NNN} interactions ($t_{2} \neq 0$). This emergent gap is protected by time reversal symmetry breaking; because now the term corresponding to $\sigma_{z}$, which breaks the time reversal symmetry, enters into the Hamiltonian. For example, if we take $\vec{q}=\left( 2 q/\sqrt{3},0\right)$, then the special points occurs at $\vec{K}=\left(\pm \pi/\sqrt{3},0 \right)$. The \emph{Chern} number in this case is:
\begin{equation}
  \label{eq:chern-num}
  c_{1} = \text{sgn} \left( \sin Sq \right), \quad -\frac{\sqrt{3} \pi}{2} \leq q \leq \frac{\sqrt{3} \pi}{2}.
\end{equation}
It is the first \emph{Chern} character of the corresponding $U(1)$ \emph{Bloch} bundle. Interestingly the \emph{Chern} number depends on the spin $S$ and on the modulation vector  $\vec {q}$.

Similarly, for the second case --- $t_{2} \neq 0$ and $\epsilon=0$ --- the Hamiltonian will be:
\begin{equation}
  \label{eq:eps-zero-hk}
  \mathcal{H}(\vec{k}) \sim \sum\limits_{n}^{3} \text{diag} \left[ \cos \left( \vec{k}-S\vec{q} \right) \cdot \vec{b}_{n}, \:\cos \left( \vec{k}+S\vec{q} \right) \cdot \vec{b}_{n} \right].
\end{equation}
By an appropriate change of the variable $\vec{k} \to \vec{k} \pm S\vec{q}$ on the \emph{A} and \emph{B} sub-lattices we will get:
\begin{equation}
  \label{eq:zero-chern-hk}
  \mathcal{H}(\vec{k}) \sim \cos k_{x} \cos k_{y}.
\end{equation}
This Hamiltonian is time reversal invariant, and exhibits no topological properties, as $c_{1}=0$. Physically, $\epsilon=0$ corresponds to the ferromagnetic phase with constant spin \emph{z} component. Due to the absence of the variation of the underlying spin structure the electrons don't acquire any \emph{Berry} phase during hopping. Hence there is no topological effect in this case.

\begin{figure}
     \centering
     \includegraphics[width=0.4\textwidth]{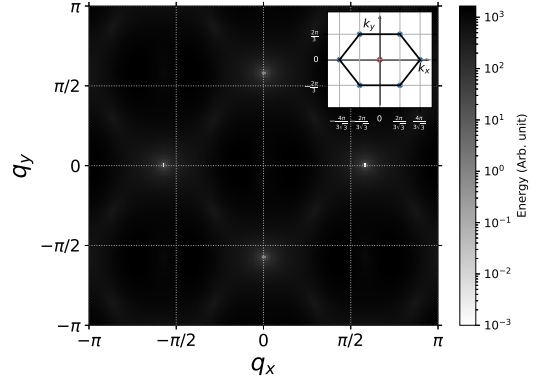}
     \caption{Dependence of the total energy of the system on the spin modulation vector $\vec{q}$ for a conical spin configuration. The total energy is found by numerically integrating Eq.~(\ref{eq:h-ener}) over the \emph{Brillouin} zone (inset). The minimum energy --- the ground state --- occurs at $\vec{q}=\left( \pm \pi/\sqrt{3},0 \right)$.}\label{fig:qmin}
   \end{figure}
The \emph{k}-th mode energy of the Hamiltonian --- Eq.~(\ref{eq:kern-hk-con-spin}) --- for the upper band is:
\begin{equation}
  \label{eq:h-ener}
  E(\vec{k}) = \mathcal{H}_{0}(\vec{k}) + \sqrt{\mathcal{H}_{x}^{2}(\vec{k}) + \mathcal{H}_{y}^{2}(\vec{k}) + \mathcal{H}_{z}^{2}(\vec{k})}.
\end{equation}
The total energy of the system is established by integrating Eq.~(\ref{eq:h-ener}) over the whole \emph{Brillouin} zone. Of course the resulting integral will be a function of $\vec{q}$ only. In Fig.~\ref{fig:qmin} we plot the full dependence of the energy of the system on $\vec{q}$. It can be observed that, for $\vec{q}=\left(\pm \pi/\sqrt{3},0 \right)$ the energy is a minimum. Hence, for the conical spin structures in the ground state $q=\pm \pi/2$.

\begin{figure}
     \centering
     \subfloat[][]{\includegraphics[width=0.22\textwidth]{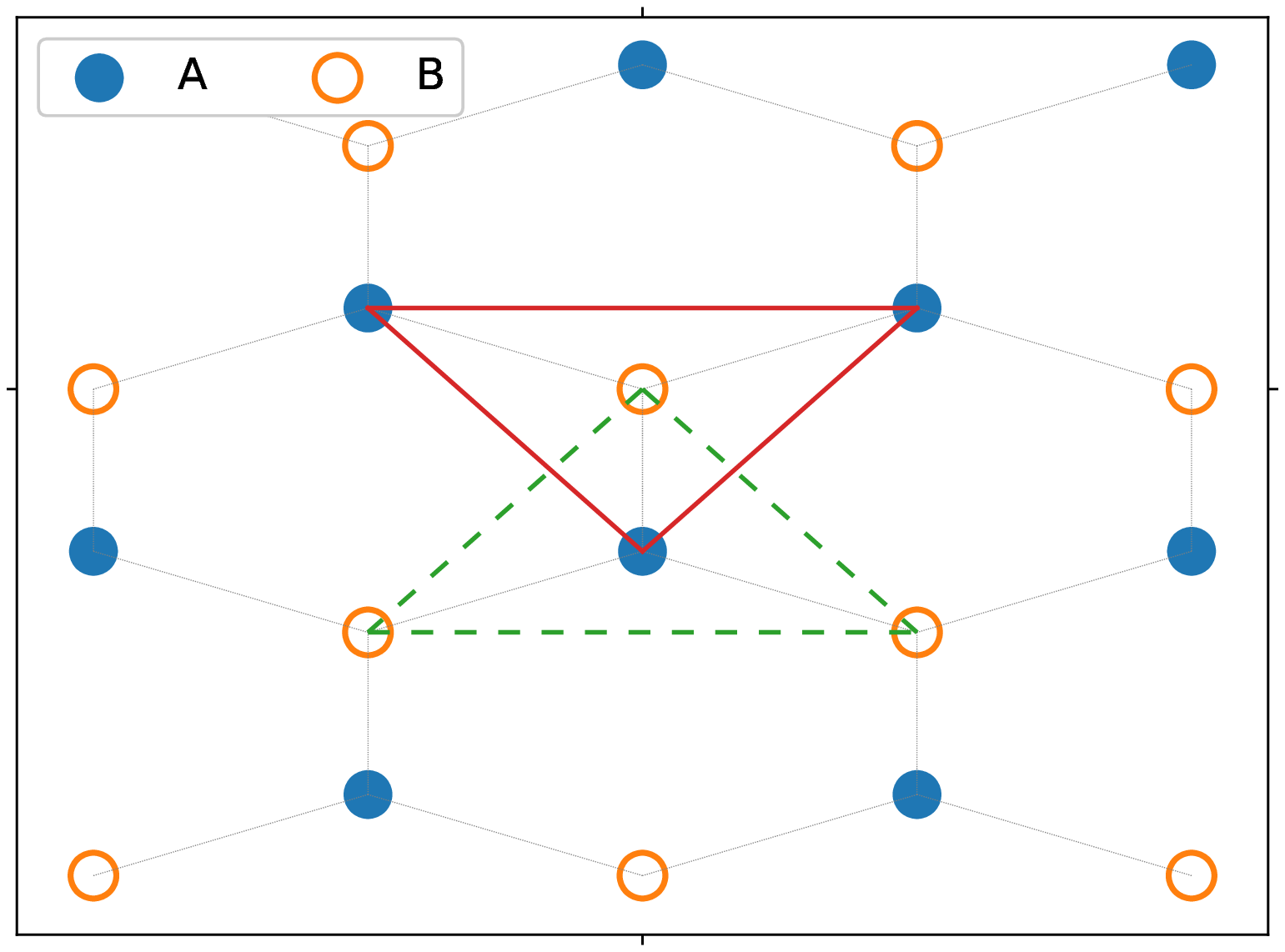}\label{fig:spin-lat}}
     \subfloat[][]{\includegraphics[width=0.22\textwidth]{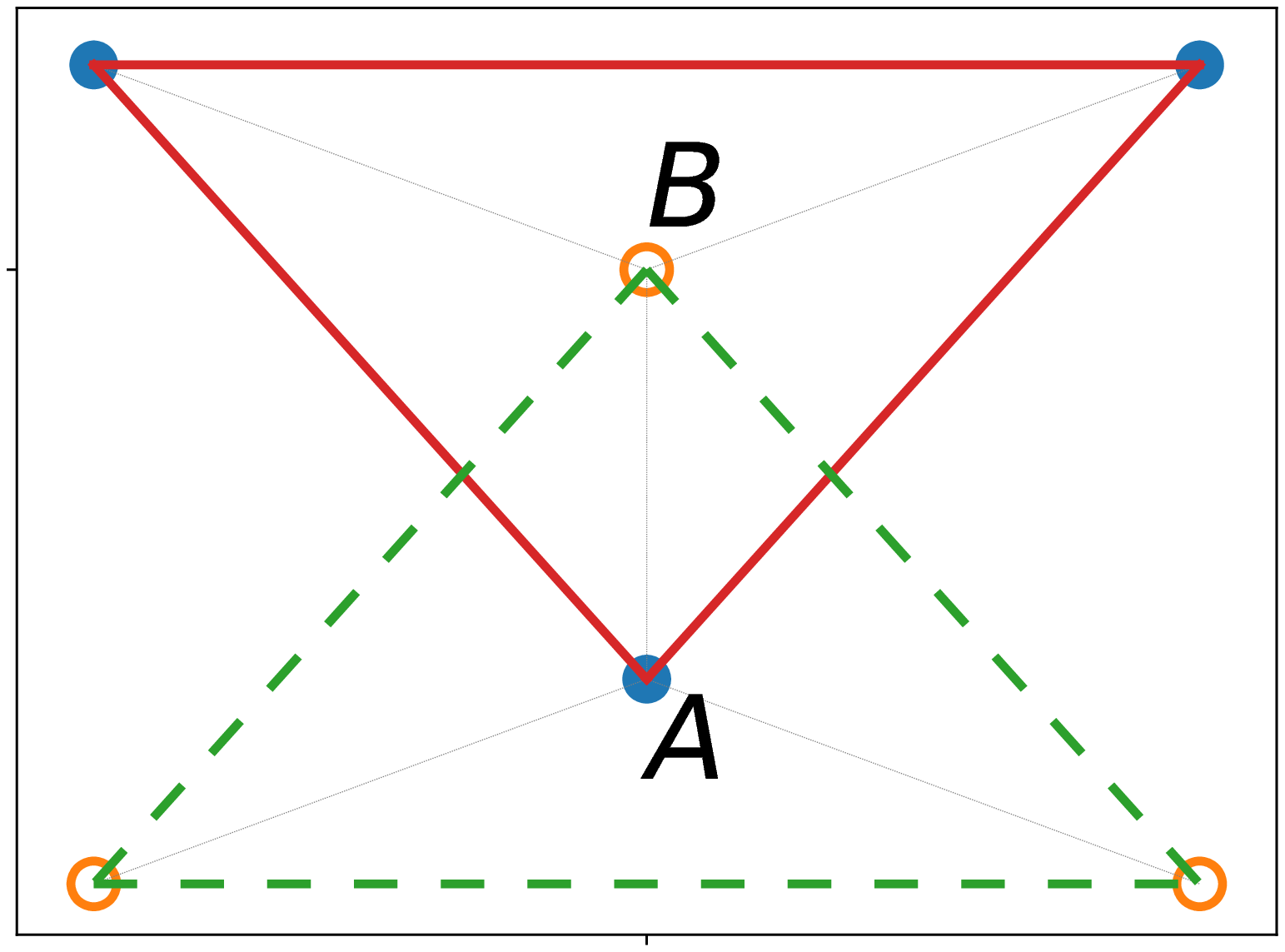}\label{fig:spin-unit-cell}}\\
     \subfloat[][]{\includegraphics[width=0.22\textwidth]{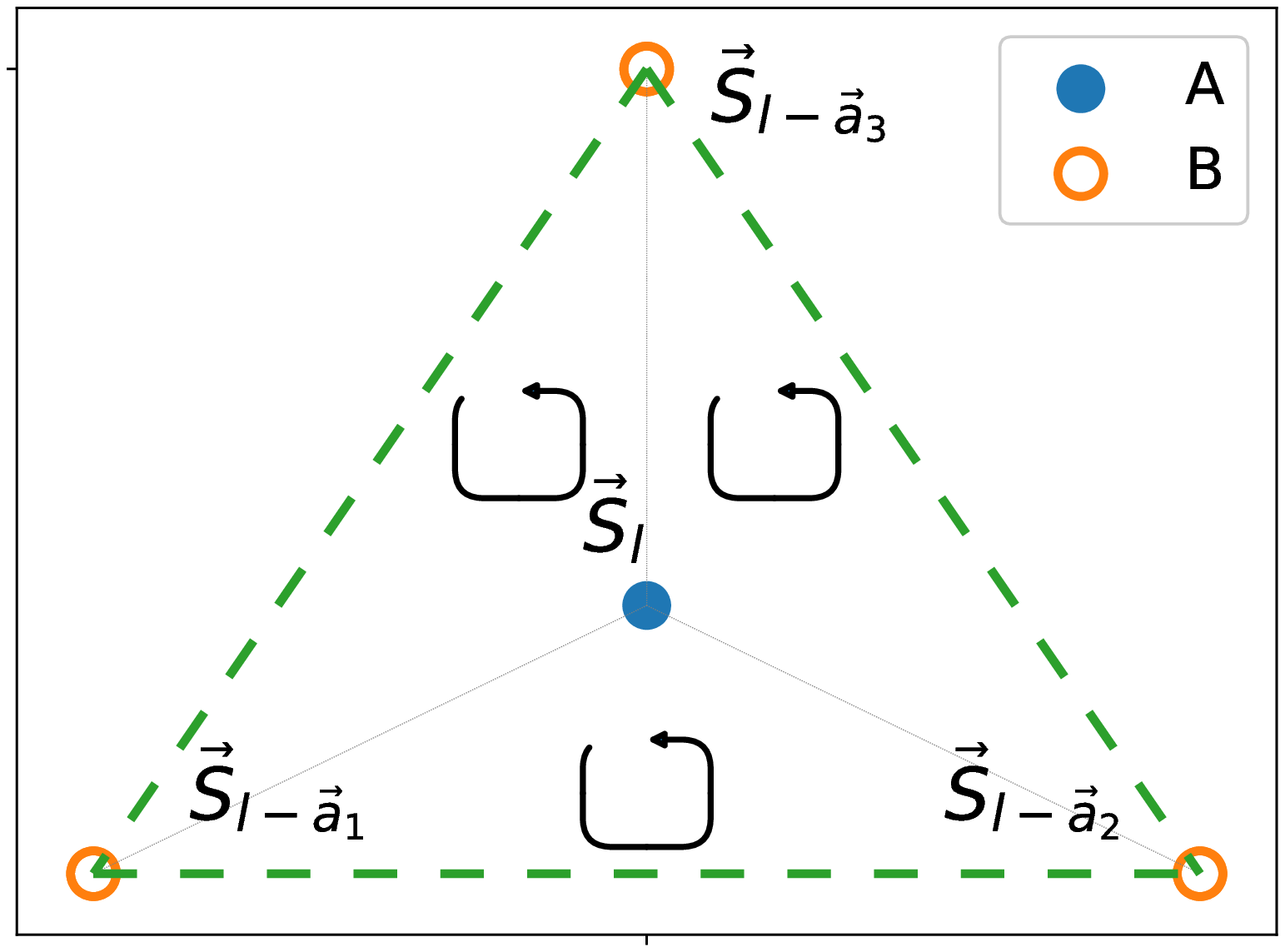}\label{fig:A-spin-lat}}
     \subfloat[][]{\includegraphics[width=0.22\textwidth]{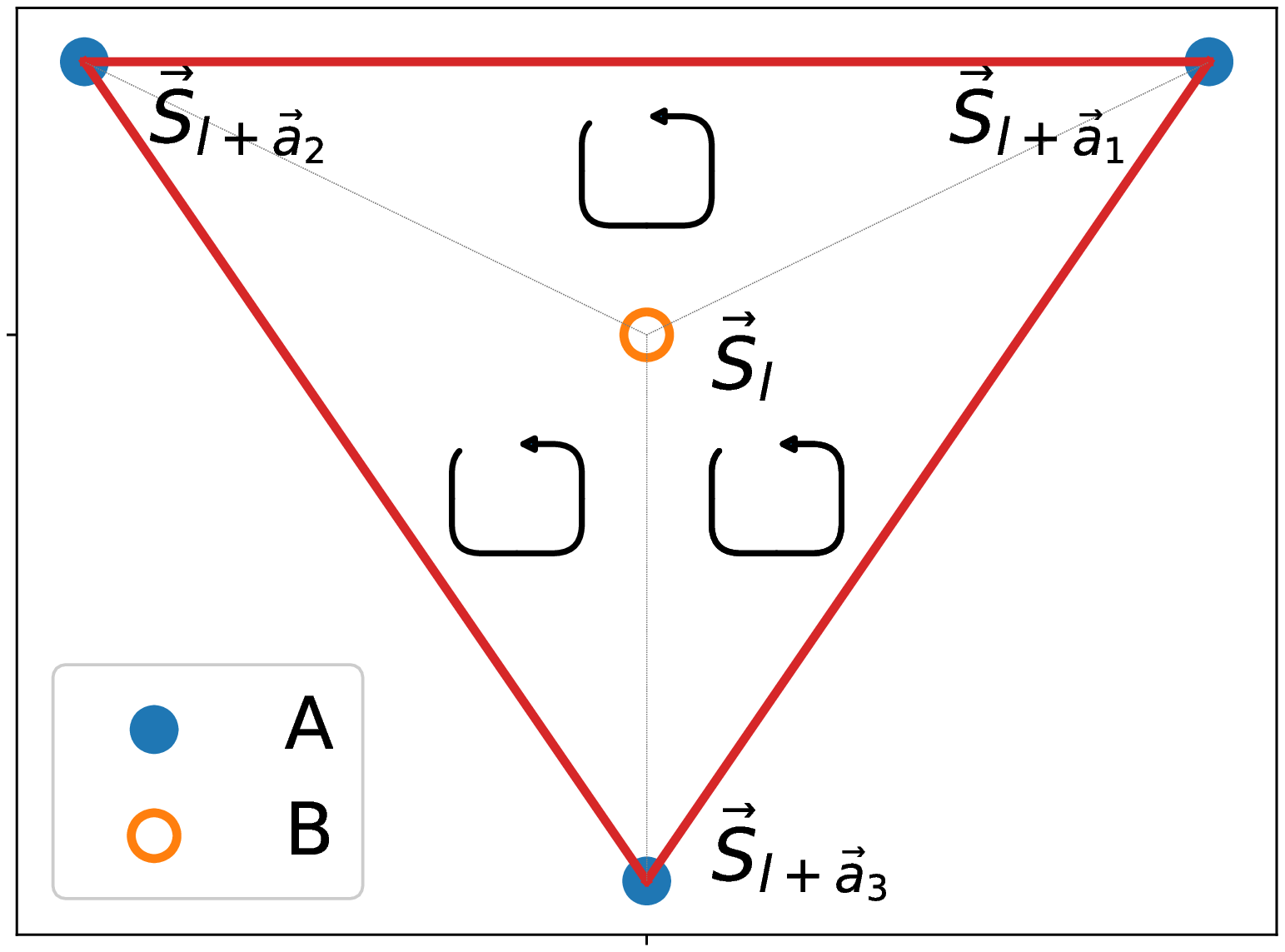}\label{fig:B-spin-lat}}\\
     \caption{(a,b) Nearest neighbours of atoms \emph{A} and \emph{B} in a honeycomb bipartite lattice. Every lattice site contains three nearest neighbours. Moreover for atom \emph{A} the nearest neighbours are atom \emph{B}, and vice-versa (c) Spin configuration of the atom \emph{A} and its nearest neighbours. $\vec{S}_{l}$ represents the spin of \emph{l}-th site. The chirality of the \emph{l}-th site is found by summing the chirality of the spin triangles: $\triangle \vec{S}_{l} \vec{S}_{l-\vec{a}_{2}} \vec{S}_{l-\vec{a}_{3}}$, $\triangle \vec{S}_{l} \vec{S}_{l-\vec{a}_{3}} \vec{S}_{l-\vec{a}_{1}}$, and $\triangle \vec{S}_{l} \vec{S}_{l-\vec{a}_{1}} \vec{S}_{l-\vec{a}_{2}}$ (d) Spin configuration of the atom \emph{B} and its nearest neighbours.}\label{fig:spin-lat}
\end{figure}
In a honeycomb bipartite lattice every site has three nearest neighbours as shown in Fig.~\ref{fig:spin-unit-cell}. Hence, the scalar spin chirality at the \emph{l}-th site for a honeycomb lattice can be calculated as:
\begin{equation}
  \label{eq:chiral-mat-rep}
  \chi_{l} = \sum\limits_{\left( m,n \right)}\vec{S}_{l} \cdot \left[ \vec{S}_{m} \times \vec{S}_{n} \right].
\end{equation}
Here $\left( m,n \right)$ represents the even permutation of the nearest neighbours of the \emph{l}-th site. The total chirality is determined by summing $\chi_{l}$ over the whole lattice. In fact for a honeycomb bipartite lattice the total chirality is the summation of the chirality of sub-lattice \emph{A} and sub-lattice \emph{B}:
\begin{equation}
  \label{eq:chiral-tot}
  \chi = \sum\limits_{l \in A} \chi_{l}^{A} + \sum\limits_{l \in B} \chi_{l}^{B}.
\end{equation}
Using Eq.~(\ref{eq:chiral-mat-rep}) we can calculate the chirality of the conical spin configuration for two adjacent \emph{A} and \emph{B} atoms connected by \emph{NN} vector $\vec{a}_{3}$, which is $\chi_{l}^{A}=-\chi_{l}^{B}$. Hence the total chirality will be $\chi=0$.

As a result we arrive at the THE in spite of the zero total chirality. This is in agreement with a number of other recent independent results which indicate the emergence of THE even if the scalar spin chirality is identically  zero~\cite{afsharSpinSpiralTopological2021,mendezCompetingMagneticStates2015,ghimireCompetingMagneticPhases2020,gongLargeTopologicalHall2021,wangFieldinducedTopologicalHall2021,daoHallEffectInduced2022,buschMicroscopicOriginAnomalous2020,chengEvidenceTopologicalHall2019,wangSpinChiralityFluctuation2019}. In all of these works the THE was ascribed to the spin fluctuation as a function of the temperature~\cite{afsharSpinSpiralTopological2021,mendezCompetingMagneticStates2015,wangSpinChiralityFluctuation2019,buschMicroscopicOriginAnomalous2020,chengEvidenceTopologicalHall2019,kimbellChallengesIdentifyingChiral2022}. At a higher temperature a small varying \emph{z}-spin component may become manifest in the conical spin configuration. In such a case the accumulated \emph{Berry} phase is non-zero and the THE emerges out of that~\cite{kimbellChallengesIdentifyingChiral2022,mendezCompetingMagneticStates2015}. However, in our model, we show that the THE state emerges even if the temperature induced fluctuations are not present in the physical system.

\section{Conclusion}
We show that the topological Hall effect can be placed into the context of the phenomena associated with strong  electron correlation. The necessary step for that is achieved by replacing the large-spin connection by its classical counterpart to  break the time-reversal symmetry. If the spin ordering opens a full gap in the charge excitation spectrum the conditions are given for the topological Hall effect to be fully manifest. We show it can arise  even in the absence of a nonzero scalar spin chirality  of the underlying spin texture. Our approach demonstrates explicitly in what way strong correlation can directly affect topology.

\begin{acknowledgments}
  K.K.K. would like to acknowledge the financial support from the JINR grant for 	young 	scientists and specialists, and RFBR Grant No. 21-52-12027. One of us (A.F.) wishes 	to acknowledge financial support from the Simons Foundation (Grant Number 1023171,RC) 	and from the Brazilian CNPq and Ministry of Education.
\end{acknowledgments}

\bibliographystyle{apsrev4-2}

\end{document}